\documentclass[doublecol]{epl2} 

\usepackage {graphicx}
\usepackage {amsmath}
\usepackage {amsfonts}
\usepackage {amssymb}
\usepackage {mathrsfs}
\usepackage {bm}

\usepackage {epstopdf}

\newcommand {\be}{\begin{eqnarray}}
\newcommand {\ee}{\end{eqnarray}}
  
\newcommand{\prb}{{Phys.\ Rev.\ B}}
\newcommand{\prl}{{Phys.\ Rev.\ Lett.}}

\title{Multiband magnetism and superconductivity in Fe-based compounds}

\author{V.\ Cvetkovic\inst{1} \and Z.\ Tesanovic\inst{1} }

\institute{                    
  \inst{1} Department of Physics and Astronomy, The Johns Hopkins University,
Baltimore, MD 21218
}

\abstract{
Recent discovery of high T$_c$ superconductivity in
Fe-based compounds may have opened a new
pathway to the room temperature superconductivity.
The new materials feature FeAs layers instead of the
signature CuO$_2$ planes of much-studied cuprates.
A model Hamiltonian describing FeAs layers is introduced, 
highlighting the crucial role of puckering of As atoms in promoting d-electron
itinerancy and warding off large local-moment magnetism of Fe ions,
the main enemy of superconductivity. 
Quantum many-particle effects in charge, spin and multiband channels
are explored and a nesting-induced spin density-wave
order is found in the parent compund. We argue that this largely 
itinerant antiferromagnetism and high T$_c$ itself
are essentially tied to the multiband nature of the Fermi surface.}

\begin{document}

\maketitle

Recently, a surprising new path to room-temperature 
superconductivity might have been discovered.
The quaternary compound LaOFeP
was already known to become superconducting
below 7K \cite {LaOFeP}, when its 
doped sibling LaO$_{1-x}$F$_{x}$FeAs ($x > 0.1$)
turned out to have unexpectedly high 
$T_c$ of 26K \cite {LaOFeAs}.
Even higher $T_c$'s were found by replacing 
La with other rare-earths (RE), reaching
the current record of $T_c =$ 55K \cite {recordTc}. 
These are the first
non-cuprate superconductors exhibiting such high $T_c$'s.

\begin{figure} 
\raisebox {1.3in} {a)}
\includegraphics[width=0.2\textwidth]{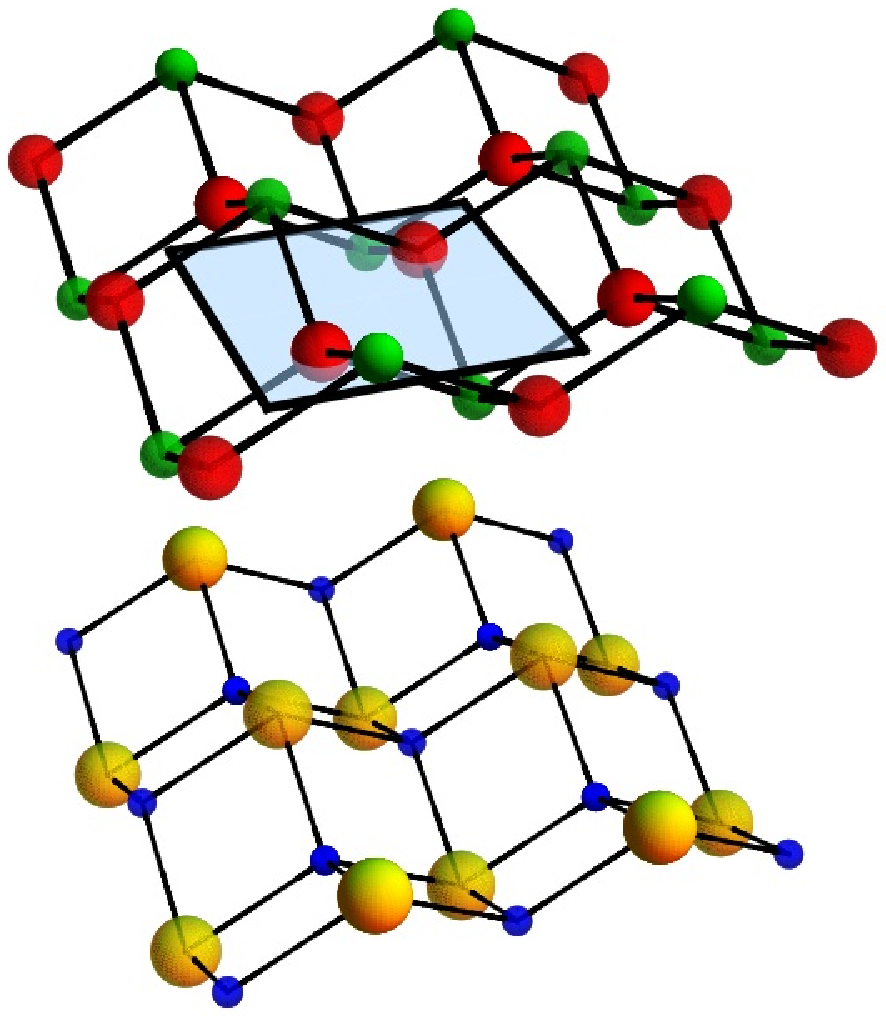}
\hspace {0.02\textwidth}
\raisebox {1.3in} {b)}
\includegraphics[width=0.2\textwidth]{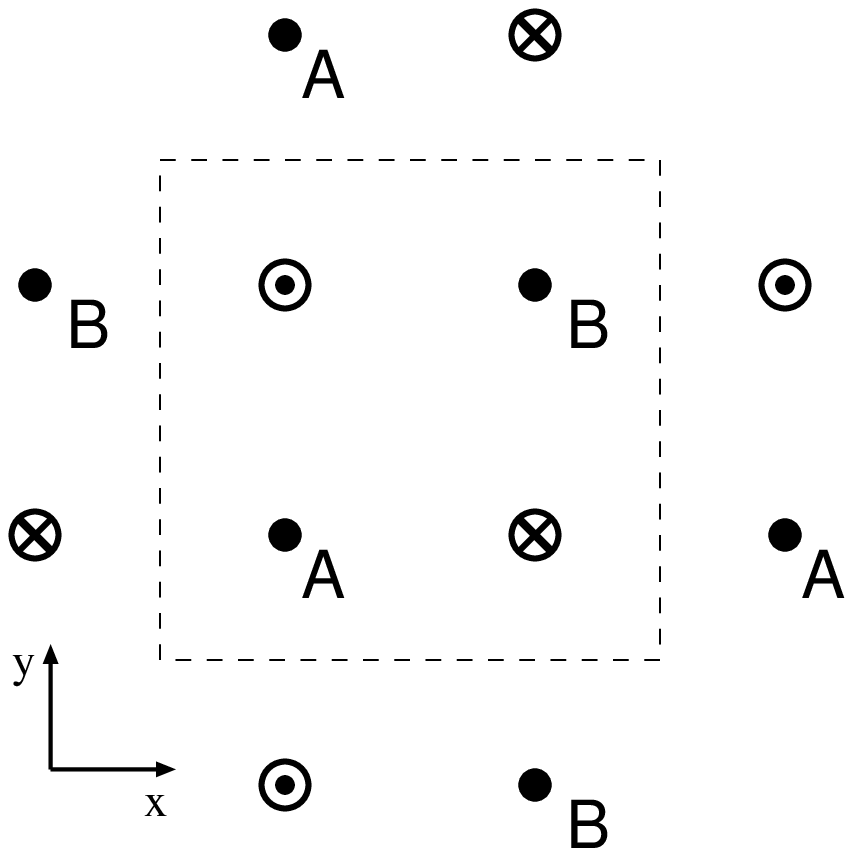}
\raisebox {1.2in} {c)}
\includegraphics[width=0.45\textwidth]{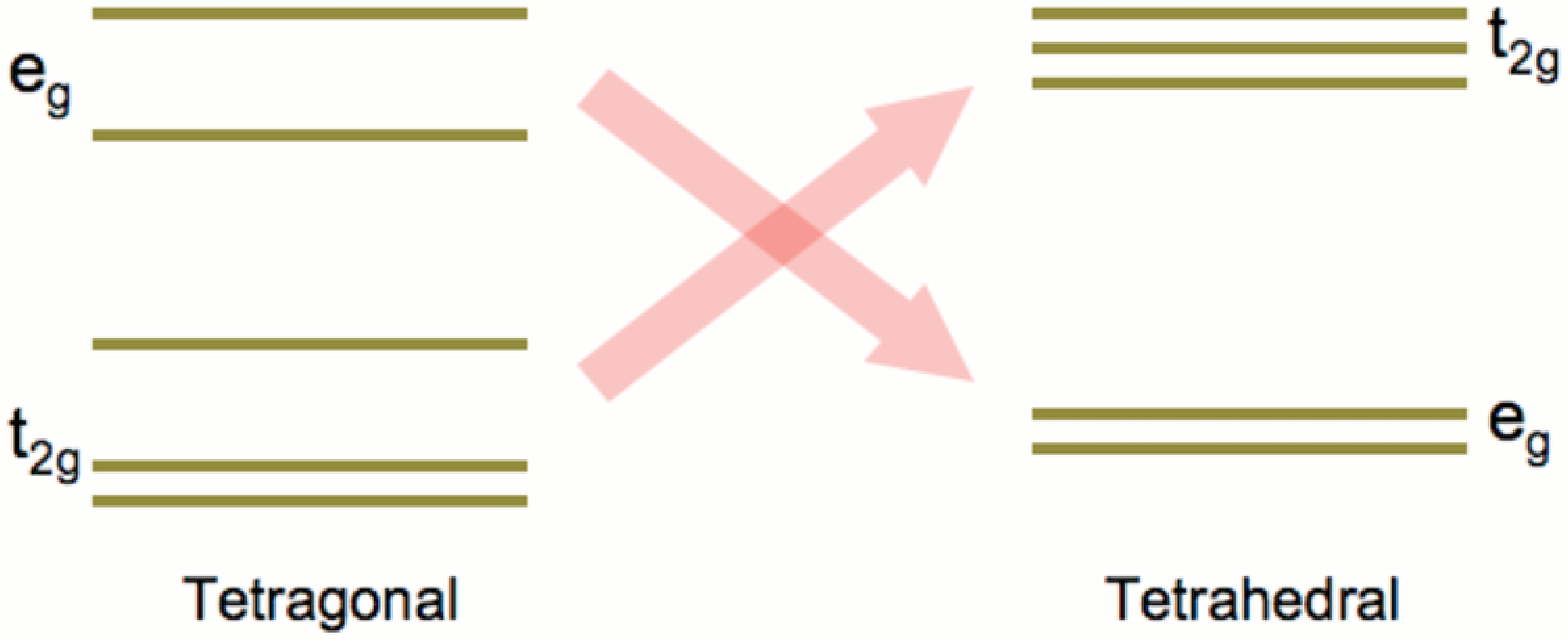}
\caption{ (Colour on-line)
(a) The three dimensional structure of the parent compound,
with FeAs layer (Fe -- red, As -- green) on top of a REO
layers (RE -- yellow, O -- blue); The blue square in the FeAs plane
corresponds to the `planar' unit cell (b).
We denote two Fe atoms with A and B, while the two As atoms that are displaced
up and down with respect to the layer are presented by dotted and crossed circles
respectively. We give our choice of axes in the corner 
(note that some papers use
a coordinate system rotated by 45 degrees).
(c) The evolution of d-orbital
energy levels from the tetragonal to tetrahedral crystal field
environment. The puckering of FeAs planes results in the
situation which is ``in between'', placing all d-orbitals near the Fermi level.}
\label {Fig_UnitCell}
\end{figure}

The surprise here
is that the most prominent characteristic of iron is its natural magnetism.
By conventional wisdom, the high $T_c$
superconductivity in RE-OFeAs 
compounds is unexpected, all the more so since
the superconductivity apparently resides in FeAs layers. 
Following standard ionic accounting, rare-earths are 3$^+$, giving away
three electrons, while As and 
O are 3$^-$ and 2$^-$, respectively. One then expects Fe 
to be in its 2$^+$ configuration,
two of its 4s electrons given away to fill As and O p-orbitals, with
assistance from a single rare-earth atom.
The remaining six d-electrons fill
atomic orbitals of Fe in the overall tetragonal As/O environment
of Fig. \ref{Fig_UnitCell}; the lower three
$t_{2g}$ orbitals should be filled while the upper two $e_g$ orbitals 
should be empty. However, the Coulomb interactions intervene via the Hund's
rule: the total energy can be reduced by making the spin part
of the atomic wavefunction most symmetric and consequently the orbital part
of it as antisymmetric as possible, reducing thereby the 
cost of Coulomb repulsion. The simplest realization of this
is to occupy a low $t_{2g}$ orbital with one spin-up and one spin-down
electron while storing the remaining four electrons into the spin-up
states. The result is a total spin
$S=2$ of Fe$^{++}$, with the associated local magnetic moment
and likely magnetism in the parent compounds. This is the situation
similar to manganese, the Fe's nearest relative, whose five d-electrons
feel the full brunt of the Hund's rule and typically line up into
a large spin state, and very different from copper, where d-orbitals are
either fully occupied or contain only a single d-hole, as in the
parent state of cuprate superconductors.
All told, the circumstances are hardly hospitable to any superconductivity,
let alone a high temperature one.

In this paper, we first argue that 
the above Hund's rule route to large local moment 
magnetism is derailed by significant
banding effects, promoting enhanced itinerancy for most Fe d-electrons. 
We show how such itinerancy arises from the combination
of two factors: a sizeable overlap among Fe and As atomic 
orbitals and the distortion of the overall tetragonal structure into 
a locally near-tetrahedral environment for Fe ions, {\em both} generated
by the {\em crucial} ``puckering'' of As atoms out of the 
FeAs planes (Fig. \ref{Fig_UnitCell}).
The puckering rearranges the $t_{2g}$ and $e_g$
crystal-field levels so that $E_{t_{2g}}\sim E_{e_g}$
-- the situation ``in between'' the purely 
tetragonal ($E_{t_{2g}}<E_{e_g}$) and
the purely tetrahedral ($E_{t_{2g}}>E_{e_g}$) --
thus bringing {\em all} d-orbitals into a close proximity 
of the Fermi level $E_F$, {\em and} maximizes direct overlap between
Fe d- and As p-orbitals.
The end result are numerous bands crossing $E_F$ and a multiply
connected Fermi surface, containing both electron and hole sections. 
We introduce a two-dimensional tight-binding model 
which captures the relevant features
of this multiband problem. Next, we argue that large number of
broad bands and the absence of large local Fe moments betrays not
only the failure of the atomic Hund's rule but, via the enhanced
metallic screening, the {\em absence of strong local correlations in general}. 
This implies the key role for the nesting properties 
and we present an {\em analytic} calculation of various responses for 
circular and elliptical bands relevant to this multiband problem.
These responses allow us to
account for the observed weak antiferromagnetic
ordering in parent materials and provide strong clues about
the superconducting mechanism itself.
In this sense, the Fe-based high $T_c$ superconductors differ
from the {\em hole}-doped cuprates and are likely 
more closely related to either
the {\em electron}-doped cuprates or other weakly to moderately
correlated superconductors.

The parent compound of the Fe-based superconductors has a ZrCuSiAs type
structure \cite {ZrCuSiAs}, with eight atoms per unit cell,
depicted in Fig. \ref{Fig_UnitCell}. 
The Fe atoms lie in a plane, same as
O atoms precisely above them, in the adjacent REO layer.
In contrast, the RE and As atoms (also located above each other)
are puckered out of plane in a checkerboard fashion. 
The amount of puckering is significant, 
turning the in-plane tetragonal structure 
in the physically relevant
FeAs layer into the nearly-tetrahedral one 
(the angle of the FeAs bond with respect to
the vertical is 58.8$^{\circ}$ as compared 
to 54.7$^{\circ}$ for a tetrahedron \cite {angles}).
As stated above, this has important consequences 
for promoting banding and rich 
orbital content near $E_F$.

There available electronic structure calculations of LaOFeP
\cite {Lebegue}, and of LaOFeAs, doped and undoped \cite {Kotliar, Mazin},
consistently convey the key information that
all five Fe 3d bands of are located at the Fermi level,
in stark contrast with the cuprates. These bands are  
hybridized with 4p orbitals of As/P located
far below the Fermi level, centered between 6 and 2eV below $E_F$.
There are five sections of the Fermi surface: 
two hole concentric, near-circular quasi-2D ones around the $\Gamma$ point,
two electron elliptical ones, centered around 
the $M$ point, and a 3D
hole band with a spherical Fermi surface around the $Z$ point. 
Given the fact
that the last one vanishes upon doping \cite {Mazin}, 
and that the relevant physics
appear to be two-dimensional, we will ignore this
Fermi surface and neglect the interlayer couplings altogether. 
This idea is used in
other works which aim to recreate the band-structure, 
either with all ten bands
\cite {Kuroki, Hirschfeld}, or with some simpler 
minimal model \cite {Raghu, Eremin}.

To illustrate the key role of the puckering of
As atoms on the electronic structure of FeAs,
let us first consider the hypothetical situation in which
all As atoms are planar (Fig.\ \ref{Fig_UnitCell}). The tetragonal
crystal field splitting pushes 3d $t_{2g}$ orbitals ($xy$, $xz$, and $yz$) 
below the $e_g$ orbitals. In this arrangement, the overlap 
of Fe $t_{2g}$ orbitals with the neighbouring As
p-orbitals either vanishes by symmetry or is very small, 
the {\em only} source of  broadening for these bands
being the direct overlap of two d-orbitals on neighbouring Fe.
The $e_g$ bands, on the other hand, do directly couple to
the 4p-orbitals of As, but, since these p-orbitals are deep 
below the Fermi level, this coupling only
pushes the $e$ bands further up, increasing the crystal field
gap. The consequence is that such
a material should likely be a band insulator, turning 
into a local moment magnet once the Coulomb effects and the Hund's
rule are turned on.
A sizeable puckering changes the situation dramatically:
first, the Fe crystal field environment turns to near-tetrahedral
instead of the tetragonal. In the purely tetrahedral case, the $t_{2g}$ 
orbitals ($xz$, $yz$, and $x^2-y^2$) reverse their
position and are {\em above} the $e_g$ levels
($xy$ and $z^2$). In the nearly-tetrahedral case of real FeAs
layers, the $t_{2g}$ orbitals are slightly above $e_{2g}$,
and the overlap due to the band formation
makes {\em all} five bands important. 
This banding is the other crucial consequence of
the puckering: since the $p_{x,y}$
orbitals are not entirely in the Fe plane, 
the overlap between these orbitals and
$xz$, and $yz$ d-orbitals becomes significant, and 
it actually provides the dominant contribution to electron/hole hopping. 
At the same time, the hopping between $x^2-y^2$ and $p$, or $z^2$ and $p$
orbitals is only slightly reduced.

Based on the above arguments, 
we construct a tight-binding model which incorporates
the hoppings to the nearest neighbours and
includes the relevant overlap integrals. This
model which reflects the key qualitative features of the 
electronic structure calculations \cite {Kotliar, Mazin}, 
and which can serve as the realistic platform 
for further {\em analytic} calculations.
As shown below, even such a simplified model must include {\em all}
five $d$ bands and is defined by the tight-binding Hamiltonian
\be
  H &=& H_0 + H_t + H_{int}, \label {Htb} \\
  H_0 &=& \sum_{i, \alpha} \epsilon^{(\alpha)} d_i^{(\alpha) \dagger} d_i^{(\alpha)}
  + \sum_{j, \beta} \epsilon^{(\beta)} p_j^{(\beta) \dagger} p_j^{(\beta)}, \label {Htb0} \\
  H_t &=& \sum_{i, j, \alpha, \beta} t_{(\alpha, \beta)} d_i^{(\alpha) \dagger} p_j^{(\beta)}
  + \sum_{i, i', \alpha, \alpha'} t_{(\alpha, \alpha')}^{Fe} d_i^{(\alpha) \dagger} d_{i'}^{(\alpha')} +
  \nonumber \\
  &&\sum_{j, j', \beta, \beta'} t_{(\beta, \beta')}^{As} p_j^{(\beta) \dagger} p_{j'}^{(\beta')} + h.c.,
  \label {Htbt}
\ee
where $H_0$ describes local 3d and 4p orbitals, and $H_t$
accounts for Fe-As, Fe-Fe, and As-As hopping, in that order. 
$H_{int}$ is the interaction term and will be discussed shortly.
The operator $d_i^{(\alpha)}$ annihilates an electron in orbital 
$\alpha$ on Fe site $i$, and analogously, 
$p_j^{(\beta)}$ annihilates an electron on site
$j$ in orbital $\beta$. The summation over 
$\alpha$ takes into account all five Fe 3d orbitals,
but due to the doubling of the unit cell, 
there are actually ten of those bands, and
the summation over $\beta$ involves all three
bands $p_x$, $p_y$, and $p_z$.

The symmetry provides important guidance in understanding
the band structure of (\ref{Htb}). For example,
at $\Gamma$ point there are two doubly degenerate bands. 
One of these must be a symmetric 
combination -- relative to A and B sites -- of $d_{xz}$, and $d_{yz}$
orbitals weakly hybridized with the As 4p bands, while the  
other is the antisymmetric combination. 
The splitting between these two dublets originates 
both from the direct spread of
the $d_{xz/yz}$ bands ($t_{xz, xz}^{Fe}$), and from 
the $p$ bands spread ($t_{p}^{As}$).
At the $M$ point, these orbitals again two degenerate
dublets, albeit in a different linear combination, which are split only by the
amount proportional to $t_{(xz, xz)}^{Fe}$. From such analyses, we 
find the orbital energies and hoppings (all in eV's)

{
\addtolength{\tabcolsep}{-1pt}
\noindent
\begin {tabular} {c | c c c c}
  $\alpha$ & $x^2-y^2$ & $z^2$ & $xy$ & $xz$ \\
  \hline
  $\epsilon_\alpha$ & -0.85 & -1.4 & -1.1 & -1.15 \\
  $t_{\alpha,\alpha}^{Fe}$ & -0.55 & -0.5 & -1.6 & -0.55 \\
  $t_{\alpha, x/y}$ & 0.65 & -1.4 & 1.5 & 3.2 \\
  $t_{\alpha, z}$ & 2.1 & 1.25 &  & 0.7 \\
\end {tabular}
\begin {tabular} {c | c c}
  $\beta$ & $x$ & $z$ \\
  \hline
  $\epsilon_\beta$ & -4.0 & -4.0 \\
  $t_{\beta, \beta}^{As}$ & -0.8 & -0.45 \\
\end {tabular}
}

The interband couplings are not tabulated and their values are
$t_{z^2, xy}^{Fe} = 0.1$, $t_{xz, yz}^{Fe} = -0.75$, and $t_{x, y}^{As} = 0.8$.

The levels $\epsilon_\alpha$ reflect our previous discussion of the
crystal field splitting: $E_{t_{2g}}\sim E_{e_g}$ on the scale of
$t$'s. The hoppings reveal that
the puckering of As atoms promoted $d_{xz/yz}$ bands to the physically most 
relevant ones, their coupling to the 4p orbitals being the strongest. 
These bands provide dominant content 
of the electronic states at $E_F$,
where they get mixed with the other states, chiefly $d_{xy}$, to 
finally form the two hole Fermi surfaces. 
Clearly, these effects are difficult to reproduce
within a simple two-band model.
In addition, significant mixing
of different d-orbitals with opposite parity relative to the FeAs planes
further reinforces the need to include the full d-orbital manifold into
the basic description.

\begin{figure} 
\raisebox {1.3in}{a)}
\includegraphics[width=0.21\textwidth]{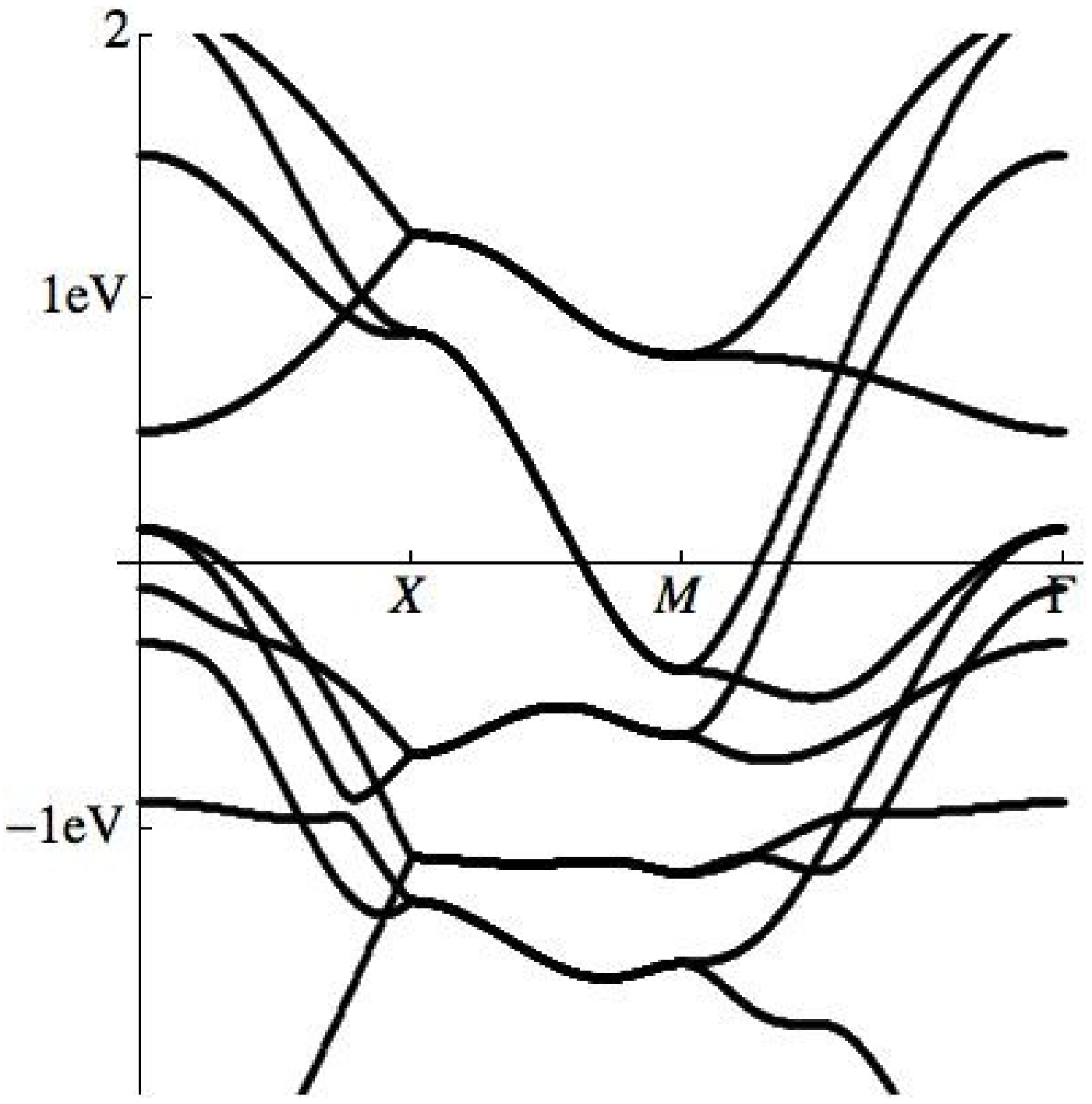}
\raisebox {1.3in}{b)}
\includegraphics[width=0.20\textwidth]{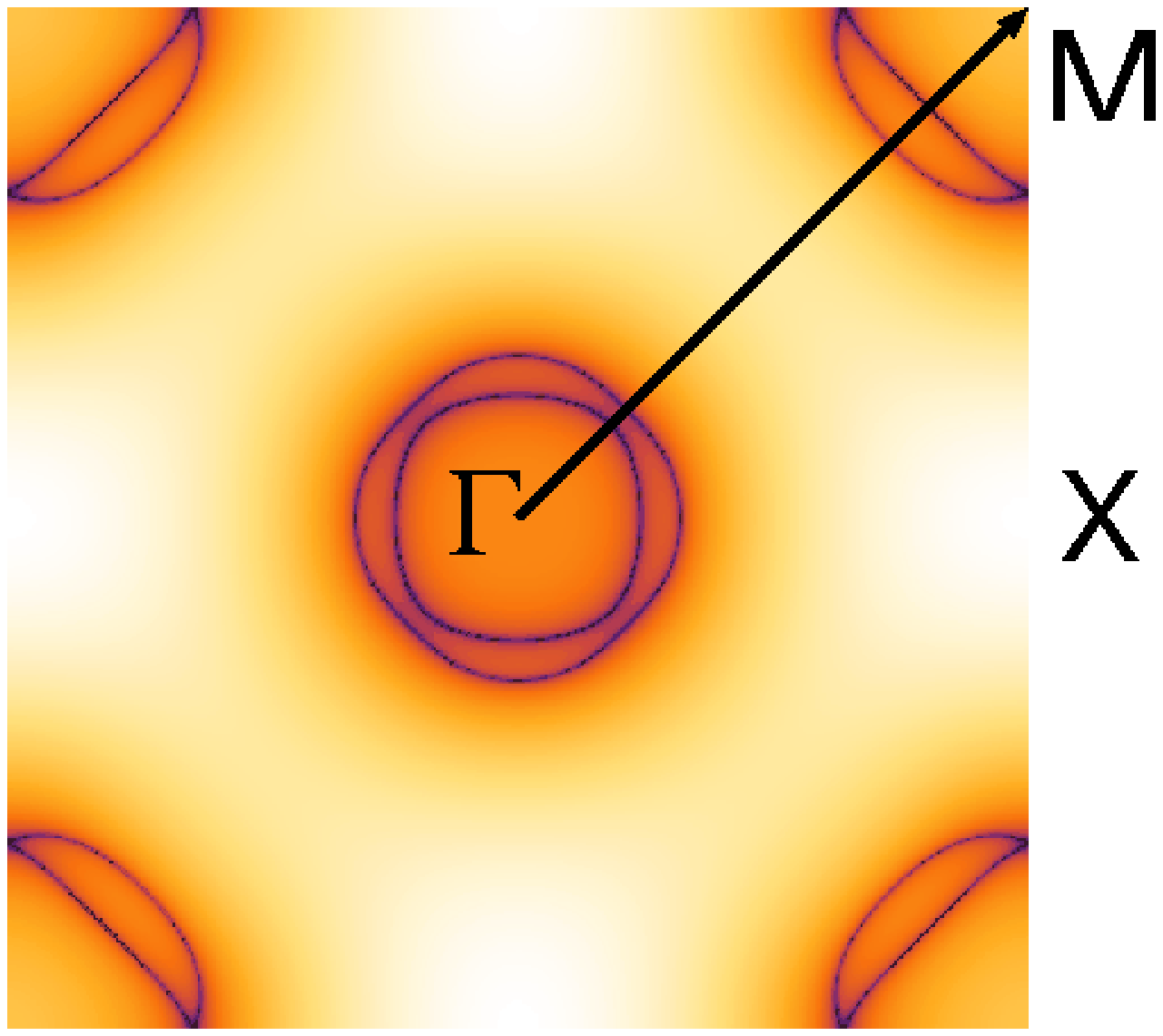}
\caption{ (Colour on-line) The band structure (a), and the Fermi surfaces (b)
following from the tight-binding Hamiltonian 
\eqref {Htb}, and using the parameters of  
the tight binding fit.}
\label {FigBands}
\end{figure}

Fig.\ \ref {FigBands} shows the band structure and the Fermi surface(s)
following from \eqref {Htb}. The key features of the Fermi surface
are faithfully reproduced, with the central hole pockets nearly circular
(actually, these are two ellipses which strongly interact and avoid
crossing). In the vicinity of the 
$M$ point, the two electron pockets have elliptical
shape and do not interact at the crossing points located at the edges of the
Brillouin zone.

This brings us to $H_{int}$ in \eqref {Htb}.
The picture of puckered As atoms discussed above, promoting
the bunching of local d-levels of Fe and their large overlap with
As p-orbitals, indicates that the d-bands are near their optimum width, given
the restrictions of dealing with 3d electrons and 4p levels far
below of $E_F$. This reduces the importance of the Hund's rule
and points to the d-electron itinerancy, rather than local
atomic (ionic) correlations, as the most relevant feature.
Indeed, this is consistent with the neutron 
scattering experiments \cite {OakRidge}, observing
weak antiferromagnetism in the parent compound below 150K
instead of the large local moment magnetism expected in
the Hund's rule limit. The AF order is suppressed by doping
and ultimately vanishes in the superconducting state.
This suggests that $H_{int}$ should generically be comparable or
smaller in magnitude than $H_t$ \eqref {Htb}. For example,
in the simple single band Hubbard model, with nearly circular
(or spherical) Fermi surface, too large on-site repulsion $U$ leads
to the ferromagnetic Stoner instability, an itinerant prelude
to the local moment formation dictated by the Hund's rule. 
We thus expect that the main effects of $H_{int}$ can be understood
by a detailed analysis of enhanced spin, charge and interband
responses of the non-interacting part of $H$ \eqref {Htb}.

With this aim in mind, we observe that various pockets
of the Fermi surface depicted in Fig. \ref{FigBands} can be viewed
as radial and elliptical distortions of the same {\em idealized}
circle, two of such ideal (hole) pockets located at $\Gamma$
and two (electron) at M points. As long as such distortions are 
not too extreme, the responses in
different channels can be evaluated {\em analytically}, 
thereby greatly facilitating theoretical analysis. Where comparison
is possible, our analytic results appear to agree with numerical 
calculations in Refs. \cite {Mazin, Raghu, Eremin}.

We first look at the spin-susceptibility, and analyze how near-nesting of
the Fermi surfaces can lead to SDW order.  To do this, we have to make some
assumptions about the Fermi surfaces and separate the 
most important contributions.
While some nesting takes place {\em within} slightly 
deformed circular Fermi surfaces in Fig. \ref{FigBands}, the main
contribution to the enhanced response arises 
from similarly shaped {\em hole and electron} pockets, followed by a less
important one arising from different hole-hole and electron-electron
pockets of the Fermi surface. 
This is easily appreciated by nothing that, for our {\em idealized}
circles, the electron-hole nesting leads to a {\em divergent} contribution
to the electron-hole propagator ({\it i.e.}, an RPA bubble).
So, we concentrate on the spin-susceptibility $\chi_s({\bf q},\omega)$
where one the particle propagators corresponds  to the 
hole band at the $\Gamma$ point with a circular Fermi surface
and Fermi momentum $k_F$, and the electron band
forming  a slightly ellipcitally deformed
Fermi surface centered around ${\bf M}$ vector, as depicted in
Fig. \ref{FigBands}. The electronic
states at the Fermi level have $k_F (1 + \xi)$ momentum if parallel to the
${\bf M}$ vector, and $k_F (1 + \eta)$ if perpendicular. The dispersions are
\be
  \epsilon_{{\bf k}}^{(e)} &=& \frac 1 {2 m_e} \left \lbrack \frac {k_x^2}{(1 + \xi )^2}
  + \frac {k_y^2}{(1 + \eta)^2}
  - k_F^2 \right \rbrack, \label {Eelectron} \\
  \epsilon_{{\bf l}}^{(h)} &=& \frac {k_F^2 - l^2}{2 m_h}, \label {Ehole}
\ee
with $m_{e, h}$ being the mass of the electron/hole band. For simplicity
we assume that they have the same mass $m_e$.
Wavevector ${\bf k}$ is given 
relative to the ${\bf M}$ point in the case of the 
electron band, while ${\bf l}$ is 
defined with respect to the center of the Brilouin zone. 
Parameters $\xi$ and
$\eta$ are tied to the eccentricity of the 
Fermi ellipse as $\varepsilon =
\sqrt {| \xi - \eta | (2 + \xi + \eta)} / (1 + {\rm max} (\xi, \eta))$, and to the ratio of
states enclosed by the two Fermi 
surfaces $N_e / N_h = (1 + \xi ) (1 + \eta)$.
Below, we evaluate the particle-hole bubble due to the
near nesting of only one hole and one electron band. 
Our results are universal, generally applicable to any situation 
involving elliptical Fermi surfaces,
and particularly relevant for FeAs, where one has to sum contributions
due to nesting of each individual hole and electron band.

\begin{figure} 
\raisebox {1.3in}{a)} \includegraphics[width=0.14\textwidth]{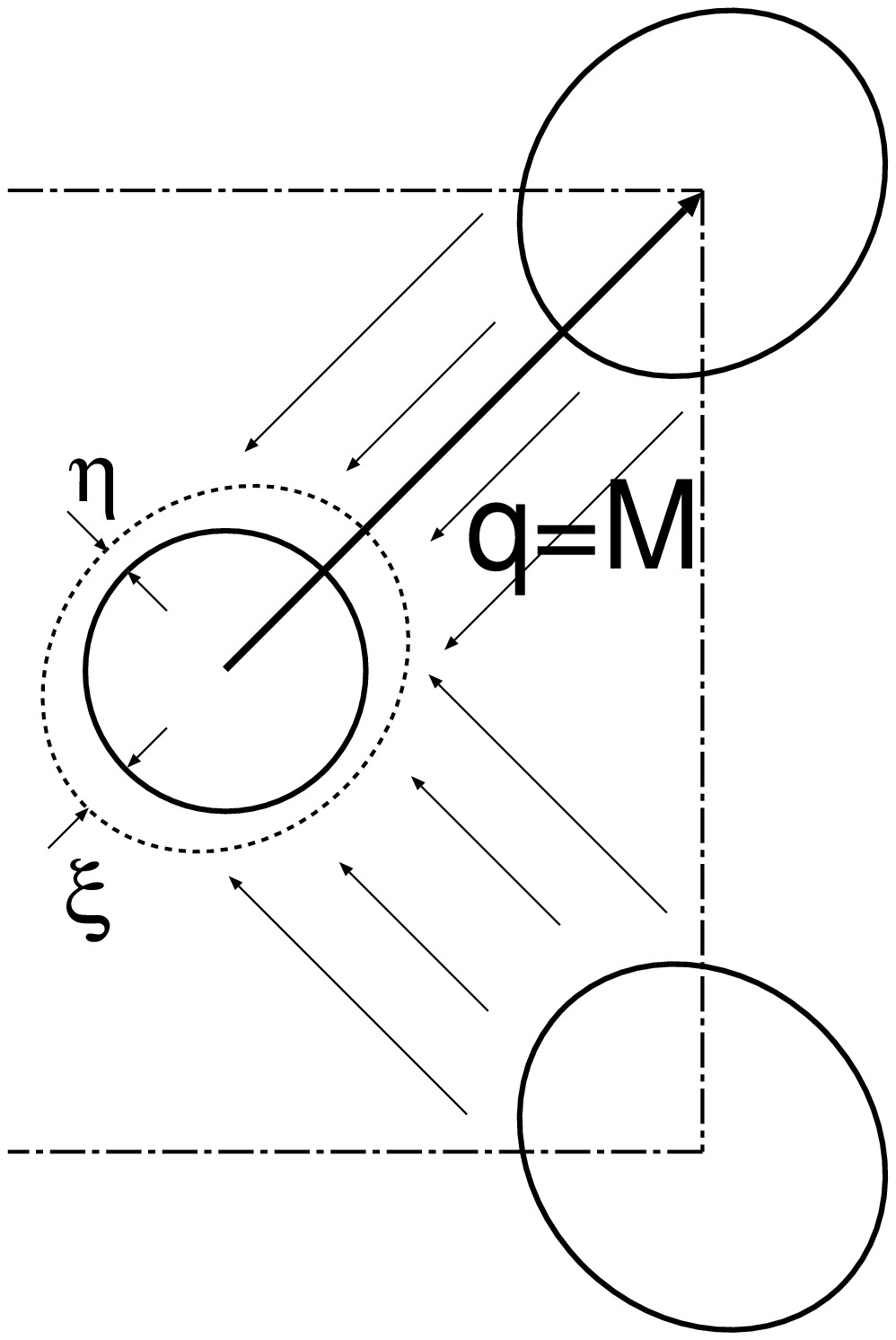}
\raisebox {1.3in}{b)} \includegraphics[width=0.24\textwidth]{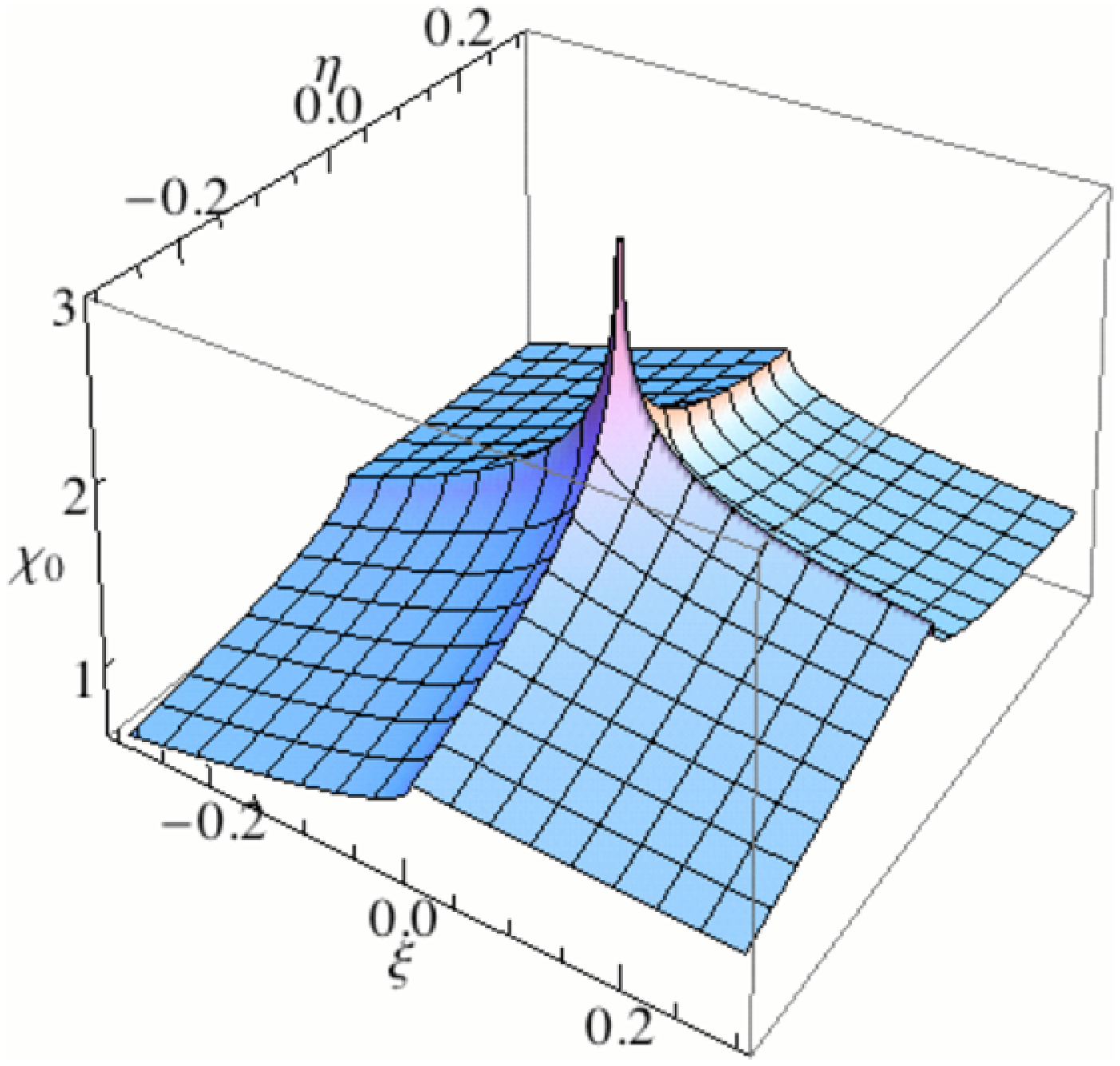}
\caption{ (Colour on-line) The arrangement of Fermi surfaces with elliptical
bands at the corners of the Brillouin Zone show in Fig. \ref{FigBands} (a), and
the regularization of the singular part of the susceptibility 
due to the elliptical distortion of the electronic Fermi surface
(b).
For $\xi = 0, \eta = 0$, the hole and the electron Fermi surfaces become
identical and the susceptibility diverges.}
\label {Fig_LogSing}
\end{figure}

If the eccentricity were zero, and the two
bands had identical Fermi momenta
($\xi = \eta = 0$), the real part of the 
susceptibility is would have simply been given by
\be
  \chi'_0 ({\bf q}, \omega = 0) =  2 \frac {m_e}{2 \pi}
  \log \frac {\Lambda} {| {\bf q} - {\bf M} |}, \label {RPAPerfect}
\ee
where $\Lambda$ is the UV band cut-off. 
A logarithmic  singularity occurrs in Eq.\ \eqref {RPAPerfect}
when ${\bf q} = {\bf M}$ due to the perfect nesting of two
hole and electron Fermi pockets. The
nesting in FeAs is not perfect due to small distortions
in Fig. \ref{FigBands}, and this singularity is cut off. Still,
it appears nevertheless that this particular response at ${\bf q} = {\bf M}$
is the strongest incipient instability of our system. If $H_{int}$
is overall repulsive and not extremely weak,
say modelled as a Hubbard repulsion on Fe sites, $U_dn_{di}^2$,
this instability will produce  the spin density-wave (SDW) ground state
at the {\em commensurate} wavevector ${\bf q} = {\bf M}$.
It seems natural to associate this Fermi surface instability
with the observed weak AFM order of the parent compound \cite {OakRidge}. 

To appreciate how the deformation of the electron
Fermi surface cuts off the singularity, we now find 
the explicit expression for this more
general situation. There are two different cases, 
depending on  whether the two Fermi surfaces
intersect after one has been moved by ${\bf M}$ 
(so that their centers coincide). If
they do not intersect (equivalent to $\xi \eta > 0$), 
the susceptibility is set by
\be
  \chi'_{0} ({\bf q} = {\bf M}, \omega = 0) = 4 \frac {m_e}{2 \pi}
  \frac {(1 + \xi) (1 + \eta)} {\sqrt { \Xi  \Upsilon } } \times \quad \qquad \label {RPAellipse_pm} \\
  \left \lbrace \log \left \lbrack \frac {\Lambda}{k_F}
  \sqrt {2 \Xi \Upsilon}  \sqrt {\Xi + \Upsilon + 2 \sqrt {\Xi \Upsilon} } \right \rbrack
  - \log ( \Xi \Upsilon) - \right . \nonumber \\
  \left . \log \left \lbrack  | \Xi \Upsilon - \Xi -\Upsilon |
  + \sqrt {\Xi \Upsilon (\Xi - 2) (\Upsilon -2)} \right \rbrack
  \right \rbrace, \quad \nonumber
\ee
where $\Xi = 1 + (1 + \xi)^2$ and $\Upsilon = 1 + (1 + \eta)^2$.

Clearly, it is the last two term which cause
the nesting divergence in the limit when 
the ellipse transforms to a circle ($\xi \to 0, \eta \to 0$). When the Fermi
surfaces do intersect ($\xi \eta < 0$), the 
last two logarithms in Eq.\ \eqref {RPAellipse_pm}
should be replaced by $- \log (2 + \xi + \eta) - \log | \Xi - \Upsilon |
$. This term is responsible for the singularity in this case. The divergent behavior 
of the real part of the susceptibility is shown in Fig.\ \ref {Fig_LogSing}.

Our analysis of the divergent part of the susceptibility was centered on
the case when ${\bf q} = {\bf M}$, and the question remains whether that is
the global maximum. The derivatives of the susceptibility with respect to
$q$ are well defined due to the regularization by finite $\xi$ or $\eta$. It is
trivial to demonstrate that the first derivative at ${\bf q} = {\bf M}$
vanishes in all directions, which can alternatively be argued based
on symmetry. Therefore, one has to look for the sign of the second
derivative in both $x$ and $y$ direction in order to determine whether the susceptibility has a
maximum, a minimum or a saddle point at ${\bf q} = {\bf M}$. Even if it
turns out that the susceptibility  has a maximum, it may be a local maximum, not the global one.
While the general treatment of the problem will be presented  elsewhere
\cite {FeAs_part2}, we illustrate the situation by
two circular Fermi surfaces with slightly different radii, $k_F$, and $k_F (1 + \xi)$.
The susceptibility due to the nesting of these Fermi surfaces is compared
for the cases when ${\bf q} = {\bf M}$, and
${\bf q} = {\bf M} + k_F \xi {\bf n}$, with ${\bf n}$ being a unit vector pointing in
an arbitrary direction. The former corresponds to concentric Fermi surfaces,
the latter to two surfaces touching each other.  The susceptibility in the former 
case follows as a special case of Eq.\ \eqref {RPAellipse_pm}
\be
  \chi'_0 ( {\bf q} = {\bf M},0) =  \frac {4 m_e}{2 \pi} \frac {(1 + \xi)^2}{\Xi}
  \log \left [ \frac {\Lambda }{k_F | \xi |} \sqrt {\frac {2 \Xi }{(2 + \xi)^2}} \right ]
  \label {chi_concentric}
\ee
and the result for touching circles is obtained by replacing the argument
under the square root by 2. This is always slightly larger than the
susceptibility following from Eq.\ \eqref {chi_concentric}, regardless the
value of $\xi$. Such a result implies a different ordering vector
$\tilde {\bf q} = {\bf M} + k_F \xi {\bf n}$, albeit only in a continuum
theory. Our system is on a lattice, and ${\bf M}$
is commensurate with the reciprocal lattice, hence any
instability at that wave-vector
will be enhanced by Umklapp processes, whereas this is not true for other
incommensurate wave-vectors such as $\tilde {\bf q}$. Furthermore,
we may argue that two Fermi surfaces touching should not produce
any unexpected divergences in the particle-hole channel, 
by simply observing the analytic expression Eq.\ \eqref
{RPAellipse_pm} when $\xi$ or $\eta = 0$.

Eq.\ \eqref{RPAellipse_pm} can be applied to all the possible pairs
of hole/electron bands found in the band structure of FeAs. There are
two circular hole surfaces of different radii paired with two electron surfaces
which are the same, except that they are rotated by 90$^{\circ}$ 
-- this just exchanges $\xi$ and $\eta$ (the unmarked 
Fermi surface in Fig.\ \ref {Fig_LogSing}a). 
For the UV cut off we choose the inverse lattice spacing.
We now compare the relative values for the doped and parent
systems, with the help from the band structure calculations.
For the undoped parent system, we estimate \cite {Lebegue} 
$\xi_1 \approx 0.27$,
$\eta_1 \approx 0.45$, $\xi_2 \approx 0.00$, and $\eta_2 \approx 0.14$, which
yields $\chi'_0\approx 5.3 m_e$ at ${\bf q}={\bf M}$. Doping
moves $E_F$ upwards, {\em increasing} the electronic, and 
{\em shrinking} the hole Fermi surfaces. The corresponding
surfaces are now further apart, so the susceptibility is 
expected to be smaller. Using Ref. \cite {Mazin}, we estimate
$\xi_1 \approx 0.72$, $\eta_1 \approx 1.11$, $\xi_2 \approx 0.35$,
and $\eta_2 \approx 0.65$, giving $\chi'_0\approx 3.8 m_e$ at $x=0.1$. 
Similar estimate is obtained by using our tight-binding band
structure of Fig. \ref{FigBands}. This is quite a bit
smaller than the undoped value, and suggests rapid suppression of our SDW
upon doping, as observed experimentally \cite{OakRidge}. 

\begin{figure} 
\raisebox {1.0in}{a)}\includegraphics[height=1.1in]{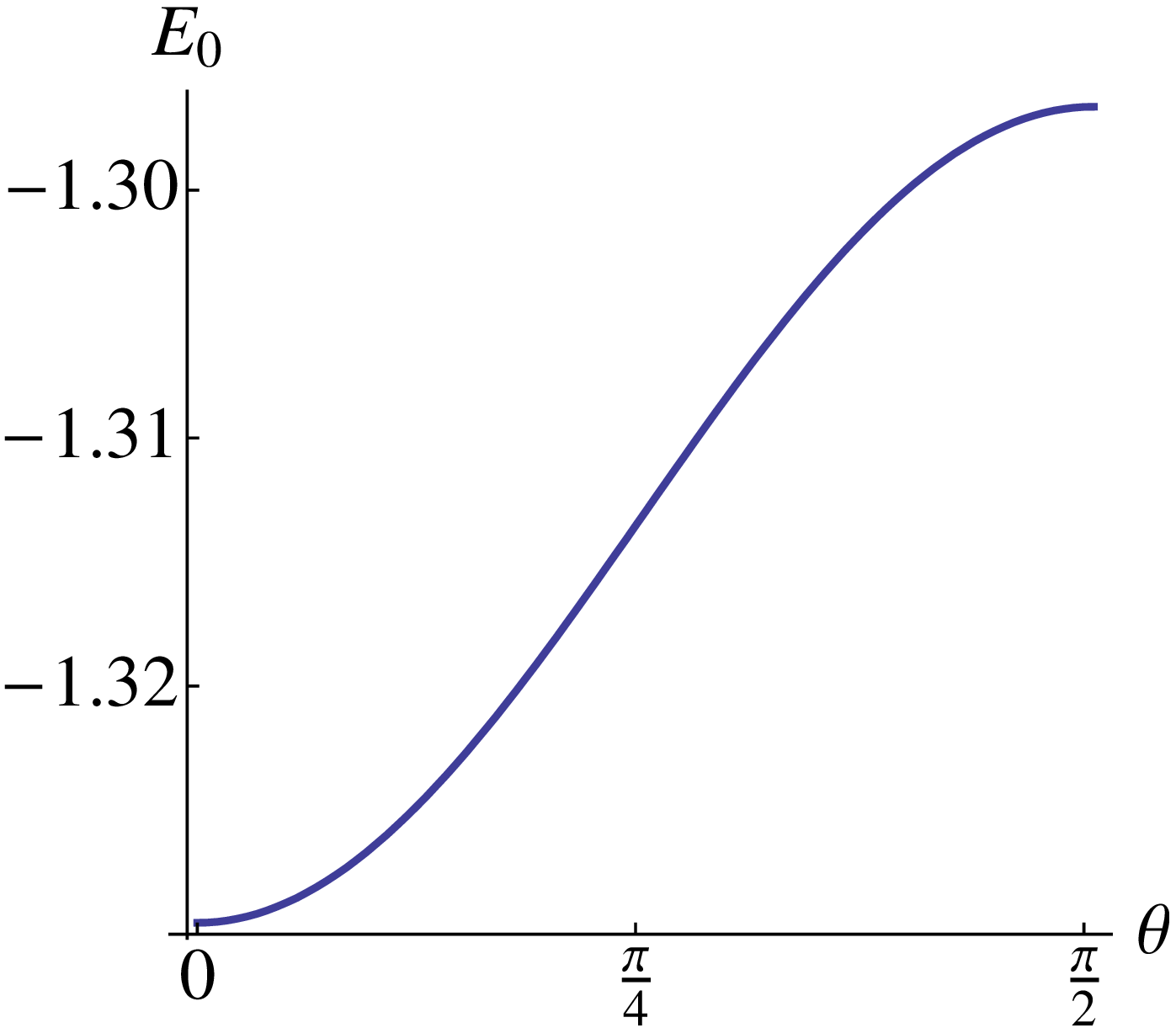} \hspace {0.2in}
\raisebox {1.0in}{b)}\includegraphics[height=1.1in]{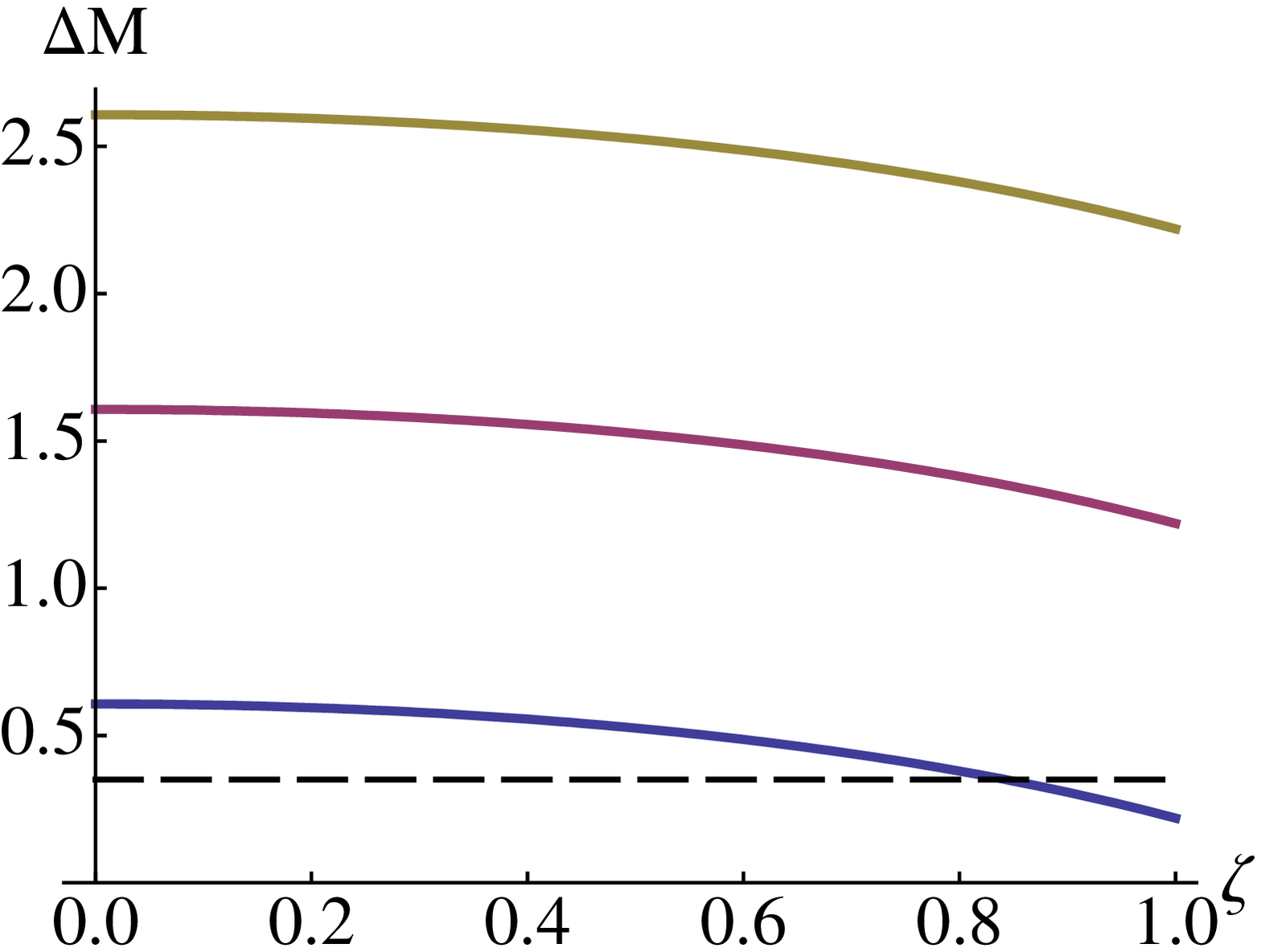}
\caption{ (Colour on-line) (a) The zero point energy in units of $J_2 S N$
as a function of angle $\theta$
between two staggered magnetizations, with
$\zeta = 0.5$ \cite {Ma}. The energy is the lowest
for $\theta = 0, \pi$. (b) The staggered magnetization in units of $\mu_B$
per unit cell is plotted for three different spin
values $S = 1/2, 1$, and $3/2$ (from bottom to top, respectively).
The dashed line corresponds
to the experimentally observed value of $0.35 \mu_B$
\cite {OakRidge}.}
\label {FigZeroPt}
\end{figure}

The SDW/AF order at ${\bf q}={\bf M}$ discussed above and
observed in experiments, could in principle also be interpreted in the {\em local}
spin picture, as arising from the direct and 
indirect superexchange between Fe atoms. 
The direct superexchange $J_1$ is generated
by the overlap of 3d orbitals of neighbouring atoms, {\it i.e.},
overlap between A and B atoms in Fig. \ref{Fig_UnitCell}b. 
Two A(B) atoms, in contrast, have an insignificant direct overlap.
However, from our band structure we
know that bands $d_{xz}$ and $d_{yz}$ hybridize with 4p orbitals of As.
Let us for example take one A atom in a unit cell in Fig.\ \ref {Fig_UnitCell},
and consider its overlap with its next neighbour A to the right. Both of
these atoms have their $d_{xz}$ orbitals hybridized with the $4p_x$ orbital
of the As atom standing between. The new 
hybridized bands both carry a significant
fraction of the As orbital, so a hopping 
between these two atoms is enabled via
the intermediate As atom. This hopping 
gives rise to the indirect superexchange
coupling $J_2$. Similar argument was
presented in Ref.\ \cite{Yildirim}.
By the same mechanism, the indirect
exchange between $d_{x^2-y^2}$ orbitals of neighboring iron atoms,
due to their overlap with p orbitals of As, yields ferromagnetic
nearest neighbor contribution to $J_1$ \cite {SiAbrahams, Manousakis}.
Our earlier analysis suggests that the
total $J_1$ is significantly smaller than $J_2$ ($\zeta= | J_1| / J_2 \ll 1$).
At such a high ratio of frustrating AF couplings, 
the AF ordering takes place individually on 
A and B sublattices \cite {PChandra} irrespective of sign of $J_1$. At the mean-field level,
the relative order on the two sublattices does not affect
the ground state energy since each B site 
interacts with four neighbouring A sites, two of these spins
pointing in the direction opposite to the other two; consequently,
there is no overall interaction. This implies that, on classical level, the
ground state would have been degenerate with its ground state energy
independent of the angle between two order parameters.
Thus, we include excitations -- spin-waves -- and investigate
how their interaction affects the ground state. For this,
we use the standard Holstein-Primakoff bosonization.
Assuming that the angle between
two order parameters on lattices A and B is $\theta$, and introducing HP
bosons $a$, and $b$ on two respective lattices, the following Hamiltonian
is obtained
\be
  \hat H &=& 2 S J_2 \sum_{\bf k} \left \lbrace 4 \label {bigH}
  (a_{\bf k}^\dagger a_{\bf k} + b_{\bf k}^\dagger b_{\bf k}) - \right .  \\
  &&(\cos k_x +\cos k_y) (a_{\bf k}^\dagger a_{-\bf k}^\dagger +
  b_{\bf k}^\dagger b_{-\bf k}^\dagger + h.c.) + \nonumber \\
  && 2 \zeta  (\cos \frac {k_x}2 \cos \frac {k_y}2) (a_{\bf k}^\dagger b_{\bf k} + b_{\bf k}^\dagger a_{\bf k}
  - a_{\bf k}^\dagger b_{-\bf k}^\dagger - a_{\bf k} b_{-\bf k}) - \nonumber \\
  &&\left . 2 \zeta \cos \theta (\sin \frac {k_x}2 \sin \frac {k_y}2)  (a_{\bf k}^\dagger b_{\bf k}
  + b_{\bf k}^\dagger a_{\bf k} + a_{\bf k}^\dagger b_{-\bf k}^\dagger + a_{\bf k} b_{-\bf k}) 
  \right \rbrace. \nonumber
\ee
The Bogoliubov transformation of the Hamiltonian \eqref {bigH} yields
two new excitations whose dispersions are given by
\begin{eqnarray}
  E_{\bf k}^\pm &=& 4 S J_2 \left \lbrack (2 + \cos k_x + \cos k_y \pm 2 \zeta \cos \frac {k_x}2 \cos \frac {k_y}2)
  \right . \times \nonumber \\
  && \left . (2 - \cos k_x - \cos k_y \pm 2 \zeta \cos \theta \sin \frac {k_x}2 \sin \frac {k_y}2)
  \right \rbrack^{\frac{1}{2}} . \label {HPdispersion}
\end{eqnarray}
We numerically determine zero point energy, and plot it in
Fig.\ \ref {FigZeroPt}(a).
The energy of the system is minimized when the spins
on two sublattices are collinear in agreement with the experiments \cite {ChandraCL}.

The staggered magnetization must be evaluated 
numerically for arbitrary $\zeta$.
Fig.\ \ref {FigZeroPt}(b) shows the magnetization
(per iron site) due to quantum fluctuations as a function of $\zeta$
for different values of spin $S$.
The fluctuations are spin independent, but the resulting magnetization
is not. Unrealistically low spin and large $\zeta$ are required 
in order to explain
what is observed experimentally \cite{OakRidge},
thus hinting at significant itinerancy in the AF state,
in line with our previous arguments.

We now turn to the superconductivity itself, clearly
the most difficult problem. It is naturally
tempting to use the above propensity for SDW at ${\bf q}={\bf M}$ in
the parent compund to generate pairing interaction once the
system is doped away from the AF order \cite{scalapino}. 
Following the example of
{\em electron}-doped cuprates and various organic superconductors, 
this would imply an ordering of a {\em nodeless} kind, 
with electron and hole pockets in Fig. \ref{FigBands}
fully gapped but with gap functions of different relative sign
(see, for example, Ref. \cite{Mazin}). 
It is important to stress here the crucial role played by the
Josephson-like interband scattering between hole and electron
Fermi surfaces in bringing about this form of superconductivity
(see further below, and also \cite{FeAs_part2}).
Since hole and electron pockets are not
identical, the gap magnitudes would not be either, but the difference could
be quite small. Observing such a relative sign difference 
in the otherwise fully gapped state (an $s\pm$ or $s'$ state) would
clearly be strong boost for this picture of superconductivity
generated entirely by antiferromagnetic (SDW) spin fluctuations.
However, the above change in sign
implies sensitivity to {\em ordinary} (non-magnetic) impurity
disorder which could severely suppress $T_c$ and the gap. 
This effect, while still present on general grounds, appears
less consequential in {\em hole}-doped cuprates, due to their strongly
correlated, almost local nature. In Fe-based superconductors the
correlations are not that strong, as we have just argued above, and
this impurity sensitivity becomes an important issue.
Finally, in order to generate the $s\pm$ (or $s'$) superconducting state,
the interband interaction -- enhanced by the proximity to the SDW -- must
overcome the {\em intraband} repulsion (see below), not an 
easy task \cite{FeAs_part2}.

In light of this, one should not out of hand dismiss the possibility that
Fe-based superconductors are of entirely different kind from even
the {\em electron}-doped cuprates and similar superconductors, where
the purely {\em repulsive} interactions suffice to generate pairing
near a magnetic instability. Their multiband
nature could instead be a realization of the exciton-assisted
superconductivity. The phonon interaction appears to be too weak to
push $T_c$ to 55 K {\em by itself} \cite{Mazin}. However, large
number of highly polarizable bands around the Fermi surface leads to
strong metallic screening and a possibility of a {\em dynamical}
overscreening, which turns $\mu$, the familiar Coulomb pseudopotential
of the Eliashberg theory, {\em attractive} in the certain {\em finite}
wavevector and frequency regions. This would pave 
the way for the exciton-assisted superconductivity, long-anticipated 
but never unambiguously observed \cite{Ginzburg}. The basic idea
is that the dynamical Coulomb interaction:
\be
V({\bf q},\omega)=\frac{V({\bf q})}{\epsilon ({\bf q},\omega)}~,
~\epsilon ({\bf q},\omega) = 1 + V({\bf q})\chi_\rho ({\bf q},\omega)~,
\label{excitons}
\ee
($V({\bf q})=4\pi e^2/q^2$)
turns attractive at some finite ${\bf q}$ and relatively low $\omega$.
Fe-based superconductors appear to have all the ingredients:
their highly polarizable multiband Fermi surface produces {\em neutral}
plasmon modes corresponding to electron and hole densities oscillating in
phase (neutral plasmons). Such modes act as ``phonons'', particularly if 
$m_e$ and $m_h$ are sufficiently different. Furthermore, the
nesting features lead to enhanced 
density response $\chi_\rho ({\bf q},\omega)$ near ${\bf q}={\bf M}$
and this could turn
the effective interaction attractive at relatively low $\omega$.
Finally, the interband pairing \cite{Suhl} could still be
essential, and acts to further boost
$T_c$ {\em irrespective} of its sign:
\be
T_c \sim\omega_p \exp
\Bigl\{
-\frac{\frac{1}{2}(\lambda_{ee}+\lambda_{hh})}{\lambda_{ee}\lambda_{hh}-\lambda_{eh}^2}+\frac{\bigl[\lambda_{eh}^2+\frac{1}{4}(\lambda_{ee} -\lambda_{hh})^2\bigr]^{\frac{1}{2}}}{\lambda_{ee}\lambda_{hh}-\lambda_{eh}^2}
\Bigr\}
\label{interband}
\ee
where $\lambda_{ee(hh)}$ and $\lambda_{eh}$ are the e-e (h-h)
and the interband coupling constants, respectively, and $\omega_p$
is the characteristic frequency of the exciton-assisted pairing.
$T_c$ generated by this mechanism is notoriously difficult to
estimate, both due to the competition from structural and covalent
instabilities in the particle-hole channel
and the need to consider local-field contributions
to $\epsilon ({\bf q},\omega)$ \cite{Littlewood}. Nevertheless, this 
``hybrid'' option -- in which
phonons help make {\em intraband} interactions attractive, or at least
only weakly repulsive, and enable the magnetically-enhanced
repulsive {\em interband} interactions provide
the crucial boost to $T_c$ -- 
should be kept in mind as the experimental and theoretical investigations
of Fe-based superconductivity continue in earnest.

In summary, Fe-based superconductors appear to offer a glimpse
of a new road toward room-temperature superconductivity.
We have constructed here a simplified tight-binding
model which qualitatively describes the physics
of FeAs layers where the superconductivity apparently resides. 
Analytical results were given for the elementary particle-hole response in
charge, spin and multiband channels and used to
discuss various features of the SDW/AF order and superconductivity.
We stress the importance of puckering of As atoms in promoting d-electron
itinerancy and argue that high $T_c$ of Fe-based superconductors
might be essentially tied to the multiband character of their Fermi surface,
favorable to the $s\pm$ (or $s'$) superconducting state.
It is tempting to speculate that different $T_c$'s obtained
for different rare-earth substitutions might be related to
the different amount of puckering in FeAs layers and regulating 
this amount might be the key to even higher $T_c$.

\section {Acknowledgments}

The authors are grateful to I. Mazin, C. Broholm and C. L. Chien  
for useful discussions. This work was supported in part 
by the NSF grant DMR-0531159 and by the DOE grant DE-FG02-08ER46544. 

\section {Additional remark}

Since this manuscript was originally posted on the arXiv in April 2008
(http://arxiv.org/abs/0804.4678) there were numerous significant 
experimental and theoretical developments in this exceptionally fast-paced field.
Those most relevant to this work include the observation of
the superconducting gap in PCAR \cite{CLC}
and ARPES experiments \cite{Ding,Wray,Kaminski}
and various theoretical approaches exemplified by 
Refs.\ \cite {vdBrink, Berkeley, Sknepnek, Sudbo, Kulic, Antropov, Podolsky}.
Especially notable among these are further theoretical explorations
of the $s\pm$ (or $s'$) superconducting state in Refs. 
\cite{Chubukov,FeAs_part2,Bernevig,Stanev}.

\bibliographystyle{apsrev}

\end {document}